\documentclass[12pt,a4paper]{article} 
\usepackage{amsfonts}
\usepackage{amsmath}

\newcommand{\mR}{{\mathbb R}}
\newtheorem{mth}{Theorem}

\title{On uncertainty relations\\ in noncommutative quantum mechanics}
\author{Katarzyna Bolonek\\ Piotr Kosi\'nski\thanks{supported by KBN grant no 5P03B05620}\\
	 Department of Theoretical Physics II\\
	 University of \L{}\'od\'z\\
	 Pomorska 149/153, 90-236 \L{}\'od\'z\\
	 Poland}
\begin{document}
\maketitle
\begin{abstract}
\small\noindent We discuss the uncertainty relations in quantum mechanics on nocommutative plane.
In particular, we show that, for a given state, at most one out of three basic nontrivial 
uncertainty relations can be saturated. We consider also in some detail the case of angular 
momentum eigenstates.
\linebreak
\end{abstract}

In recent years noncommutative field theories (NQFT) have become very popular, mainly due to 
their relation to string theory in nontrivial backgrounds \cite{1} and M--theory 
compactifications \cite{2}. As in the commutative case, the low energy limit of NQFT 
single--particle sector can be taken which results in noncommutative counterpart of 
one--particle quantum mechanics. There appeared many papers devoted to the study of
various aspects of noncommutative quantum mechanics \cite{3}-\cite{22}. Their authors 
were mainly studying the spectra of hamiltonians describing particles moving on various 
noncommutative backgrounds.

In the present letter we analyse some very simple aspects of noncommutative quantum mechanics
on the plane. Namely, we consider the uncertainty relations following from 
the basic commutation rules which define the theory. In particular, we are interested in the 
states which saturate the uncertainty relations and show that, contrary to the commutative 
case, not all basic nontrivial relations can be saturated simultaneously. Then
we concentrate on eigenstates of angular momentum and find that while $x$--$x$ relation can
be saturated exactly, $x$--$p$ ones --- only with arbitrary good but not ideal 
accuracy.

The noncommutative quantum mechanics (NCQM) on the plane is defined by the following set
of commutation relations:
\begin{subequations}
\label{w1}
\begin{align}
[\hat{x}_i,\hat{x}_j]&=i\theta\epsilon_{ij}I&\label{w1a}\\
[\hat{x}_i,\hat{p}_j]&=i\hbar\delta_{ij}I&i,j=1,2\label{w1b}\\
[\hat{p}_i,\hat{p}_j]&=0;&\label{w1c}
\end{align}
\end{subequations}
up to renumbering $1\leftrightarrow 2$ one can assume $\theta\geq 0$.

Algebra~(\ref{w1}) looks like a deformation of Heisenberg--Weyl algebra; however, both 
algebras are, in fact, equivalent: the substitution
\begin{equation}
\begin{array}{l}
\hat{x}_i\equiv\tilde{x}_i-\frac{\theta}{2\hbar}\epsilon_{ij}\tilde{p}_j\\
\\
\hat{p}_i\equiv\tilde{p}_i
\end{array}
\label{w2}
\end{equation}
transforms (\ref{w1}) into Heisenberg--Weyl algebra. The physical contents of both
theories are, in general, different and depend on the interpretation of the relevant operators.

The angular momentum operator $\hat{L}$ is defined as follows
\begin{equation}
\hat{L}\equiv\epsilon_{ij}\hat{x}_i\hat{p}_j+\frac{\theta}{2\hbar}\hat{p}_i\hat{p}_i
\label{w3}
\end{equation}
and obeys
\begin{equation}
\begin{array}{l}
[\hat{L},\hat{x}_i]=i\hbar\epsilon_{ij}\hat{x}_j\\
\\
{}[\hat{L},\hat{p}_i]=i\hbar\epsilon_{ij}\hat{p}_j
\end{array}
\label{w4}
\end{equation}
Finite rotations are represented by unitary operators
\begin{equation}
U(\alpha)=e^{\frac{i\alpha\hat{L}}{\hbar}}
\label{w5}
\end{equation}
Eqs.~(\ref{w4}) imply 
\begin{equation}
U(\alpha)
\left(
\begin{array}{c}
\hat{x}_1\\
\hat{x}_2
\end{array}
\right)U^\dagger(\alpha)=
\left(\begin{array}{rr}
\cos\alpha & -\sin\alpha\\
\sin\alpha & \cos\alpha
\end{array}\right)
\left(\begin{array}{c}
\hat{x}_1\\
\hat{x}_2
\end{array}\right)
\label{w6}
\end{equation}
together with the similar rule for $p$'s.

Commutation rules~(\ref{w1}) imply some uncertainty relations. Let us first 
remind the general scheme for such relations \cite{23} (see also \cite{24}). Given two selfadjoint operators 
$\hat{A}$, $\hat{B}$ obeying
\begin{equation}
[\hat{A},\hat{B}]=i\hat{C}
\label{w7}
\end{equation}
one can derive the following inequalities
\begin{equation}
\Delta A_\psi\cdot\Delta B_\psi\geq\frac{1}{2}|\langle\hat{C}\rangle_\psi|
\label{w8}
\end{equation}
for any normalized state $\psi$; here 
\begin{equation}
\Delta A_\psi\equiv\sqrt{\displaystyle(\psi,(\hat{A}-\langle A\rangle_\psi I)^2\psi)},
\text{\ etc.}
\label{w9}
\end{equation}
Morover, (\ref{w8}) is saturated iff
\begin{equation}
(\hat{A}-\langle A\rangle_\psi I)\psi=-i\gamma(\hat{B}-\langle B\rangle_\psi I)\psi
\label{w10}
\end{equation}
for some $\gamma\in\mR$.
Multiplying both sides of eq.~(\ref{w10}) by $\hat{A}-\langle A\rangle_\psi I$ and using 
again (\ref{w10}) as well as the saturated version of (\ref{w8}) we obtain, 
provided $\gamma\neq 0$, 
\begin{equation}
\begin{array}{l}
(\Delta A)^2_\psi=\frac{\gamma}{2}\langle\hat{C}\rangle_\psi\\
\\
(\Delta B)^2_\psi=\frac{1}{2\gamma}\langle\hat{C}\rangle_\psi
\end{array}
\label{w11}
\end{equation}
In general, the operators entering uncertainty realtions are unbounded so necessary
assumptions concerning their domains should be made.

Below we shall need yet another property: assume two selfadjoint operators $\hat{A}$ and $\hat{B}$ commute
to unit operator, $[\hat{A},\hat{B}]=icI$, $c\neq 0$; then neither $\hat{A}$ nor
$\hat{B}$ have normalizable eigenvectors (in their common invariant domain).

Consider now the uncertainty relations implied by the commutation\linebreak rules~(\ref{w1}):
\begin{subequations}
\label{w12}
\begin{align}
\Delta x_1\Delta x_2\geq\frac{1}{2}\theta\label{w12a}\\
\Delta x_1\Delta p_1\geq\frac{1}{2}\hbar\label{w12b}\\
\Delta x_2\Delta p_2\geq\frac{1}{2}\hbar\label{w12c}
\end{align}
\end{subequations}
We would like to know whether there exist states saturating the above inequalities.
In the ''classical'' ($\theta=0$) case there exist states saturating simultaneously both 
(\ref{w12b}) and (\ref{w12c}); these are famous coherent states. 
For $\theta\neq 0$ situation is different. In fact we have the following theorem:
\begin{mth}
\label{th1}
For a given state $\psi$ at most one of the uncertainty relations~(\ref{w12}) can
be saturated. 
\end{mth}
Define 
$\breve{x}_i\equiv\hat{x}_i-\langle\hat{x}_i\rangle_\psi I$,
$\breve{p}_i\equiv\hat{p}_i-\langle\hat{p}_i\rangle_\psi I$;
then $\breve{x}_i$, $\breve{p}_i$ satisfy the same algebra (\ref{w1}). Assume now
that $\psi$ saturates both (\ref{w12a}) and (\ref{w12b}). Then
\begin{equation}
\begin{array}{l}
\breve{x}_1\psi=-i\gamma \breve{x}_2\psi\\
\\
\breve{x}_1\psi=-i\delta \breve{p}_1\psi;
\end{array}
\label{w13}
\end{equation}
note that both $\gamma\neq 0$, $\delta\neq 0$; indeed, according to the remark made above
$\psi$ cannot be an eigenvector (in particular --- null eigenvector) of $\breve{x}_1$.
Eqs. ~(\ref{w13}) imply 
\begin{equation}
(\gamma\breve{x}_2-\delta\breve{p}_1)\psi=0
\label{w14}
\end{equation}
Then, due to 
\begin{align}
[\gamma\breve{x}_2-\delta\breve{p}_1,\breve{p}_2]=i\hbar\gamma I,&\ \ \ \gamma\neq 0,
\label{w15}
\end{align}
$\psi$ cannot be normalizable. This result is not very surprising: for $\theta=0$ the
coherent states for $x_1$, $p_1$ are dispersionless neither with respect to $x_1$ nor
$x_2$; however, a priori it could happen that the states saturating (\ref{w12a}) and
(\ref{w12b}) diverge in $\theta\rightarrow 0$ limit.

Assume in turn, that (\ref{w12b}) and (\ref{w12c}) are saturated for some $\psi$, i.e.
\begin{equation}
\begin{array}{lll}
\breve{x}_i\psi=-i\gamma_i\breve{p}_i\psi, & \gamma_i\in\mR, & i=1,2
\end{array}
\label{w16}
\end{equation}
Multiply both sides by $\epsilon_{ji}\breve{x}_j$ and sum over $i$, $j$
\begin{equation}
\label{w17}
\begin{split}
\theta\psi&=-i\sum_{ij}\epsilon_{ij}\breve{x}_j\breve{x}_i\psi
=-\sum_{i,j}\gamma_i\epsilon_{ji}\breve{x}_j\breve{p}_i\psi\\
&=-\sum_{i,j}\gamma_i\epsilon_{ji}\breve{p}_i\breve{x}_j\psi
=i\sum_{i,j}\gamma_j\gamma_i\epsilon_{ji}\breve{p}_i\breve{p}_j\psi=0
\end{split}
\end{equation}
i.e. $\theta=0$ or $\psi=0$. This concludes the proof of our theorem~\ref{th1}.

We are often interested in eigenvectors of $\hat{L}$; for  example, the ground 
state of rotationally invariant hamiltonian is an eigenstate of $\hat{L}$.
Assume that $\psi$ is an eigenvector of $\hat{L}$,
\begin{equation}
\hat{L}\psi=\hbar l \psi
\label{w18}
\end{equation}
i.e.
\begin{equation}
U(\alpha)\psi=e^{il\alpha}\psi
\label{w19}
\end{equation}
We have the following
\begin{mth}
\label{th2}
If $\psi$ is an eigenvector of $\hat{L}$ then neither (\ref{w12b}) nor (\ref{w12c})
are saturated.
\end{mth}
The proof is extremely simple. It follows from (\ref{w6}) and (\ref{w18})
that 
$(\psi,\hat{x}_i\psi)=(U(\pi)\psi,U(\pi)\hat{x}_i\psi)=
(U(\pi)\psi,U(\pi)\hat{x}_iU^\dagger(\pi)U(\pi)\psi)=
(\psi,U(\pi)\hat{x}_iU^\dagger(\pi)\psi)=
-(\psi,\hat{x}_i\psi)$;
on the other hand, using $U(\frac{\pi}{2})$ we find 
$(\psi,\hat{x}_1^2\psi)=(\psi,\hat{x}_2^2\psi)$; the same result holds for $p$'s.
Therefore 
$(\Delta x_1)_\psi=(\Delta x_2)_\psi$,
$(\Delta p_1)_\psi=(\Delta p_2)_\psi$ and both (\ref{w12b}) and (\ref{w12c}) must be saturated 
simultaneously. This is, however, impossible by Theorem~\ref{th1}.

Although (\ref{w12a}) and (\ref{w12b}) cannot be saturated for eigenstates of $\hat{L}$
they can be satisfied with arbitrary accuracy. First, as we mentioned above,
for eigenstates of $\hat{L}$, 
$\langle \hat{x}_i\rangle=0$,
$\langle \hat{p}_i\rangle=0$,
$\langle \hat{x}_1^2\rangle=\langle \hat{x}_2^2\rangle$,
$\langle \hat{p}_1^2\rangle=\langle \hat{p}_2^2\rangle$.
Using (\ref{w2}) we have
\begin{equation*}
\langle \hat{x}_1^2\rangle=\frac{1}{2}\langle \hat{x}_1^2+\hat{x}_2^2\rangle
=\frac{1}{2}\left\langle\frac{\theta^2}{4\hbar^2}(\tilde{p}_1^2+\tilde{p}_2^2)
+(\tilde{x}_1^2+\tilde{x}_2^2)-\frac{\theta}{\hbar}\hat{L}\right\rangle
\end{equation*}
Take null eigenvector of $\hat{L}$ and assume it saturates both 
$\Delta\tilde{x}_1\cdot\Delta\tilde{p}_1\geq\frac{\hbar}{2}$,
$\Delta\tilde{x}_2\cdot\Delta\tilde{p}_2\geq\frac{\hbar}{2}$
(this is possible because we are now dealing with standard Heisenberg operators).
Then (again
$\langle \tilde{x}_i\rangle=0$,
$\langle \tilde{x}_1^2\rangle=\langle \tilde{x}_2^2\rangle$,
$\langle \tilde{p}_i\rangle=0$,
$\langle \tilde{p}_1^2\rangle=\langle \tilde{p}_2^2\rangle$)
\begin{equation*}
\begin{split}
\langle\hat{x}_1^2\rangle&
=\frac{1}{2}\left\langle\frac{\theta^2}{4\hbar^2}(\tilde{p}_1^2+\tilde{p}_2^2)
+(\tilde{x}_1^2+\tilde{x}_2^2)\right\rangle=\\
&=\left\langle\frac{\theta^2}{4\hbar^2}\tilde{p}_1^2+\tilde{x}_1^2\right\rangle
=\frac{\theta^2}{4\hbar^2}\langle\tilde{p}_1^2\rangle+
\frac{\hbar^2}{4\langle\tilde{p}_1^2\rangle}
\end{split}
\end{equation*}
and
\begin{equation*}
\langle\hat{x}_1^2\rangle\langle\hat{p}_1^2\rangle=
\frac{\theta^2}{4\hbar^2}(\langle\tilde{p}_1^2\rangle)^2+\frac{\hbar^2}{4}
\end{equation*}
Now we can take normalizable eigenstate with arbitrary small dispersion 
$\langle\tilde{p}_1^2\rangle$; however, the state with $\langle\tilde{p}_1^2\rangle=0$
is not normalizable.

On the other hand we shall see that (\ref{w12a}) can be saturated for eigenvectors of $\hat{L}$.

We shall show now that any of the inequalities (\ref{w12}) can be saturated. Define formally 
\begin{equation}
T(\theta)=e^{\frac{i\theta}{2\hbar^2}\hat{p}_1\hat{p}_2}
\label{w20}
\end{equation}
Then $T^{-1}(\theta)=T^\dagger(\theta)=T(-\theta)$ and
\begin{equation}
\label{w21}
\begin{split}
\tilde{x}_1&=T(\theta)\hat{x}_1T^\dagger(\theta)\\
\tilde{x}_2&=T^\dagger(\theta)\hat{x}_2T(\theta)
\end{split}
\end{equation}

Let us remind again that $\tilde{x}_i$ and $\tilde{p}_i=\hat{p}_i$ form standard 
Heisenberg--Weyl algebra. Therefore, there exist coherent $|\phi\rangle$ states saturating 
\begin{equation}
\begin{array}{ll}
\Delta\tilde{x}_i\Delta\tilde{p}_i\geq\frac{\hbar}{2}, & i=1,2
\end{array}
\label{w22}
\end{equation}
Due to (\ref{w21}) we conclude then that $T(-\theta)|\phi\rangle$ ($T(\theta)|\phi\rangle$)
saturates (\ref{w12b}) ((\ref{w12c})).
It remains to show that the action of $T(\pm\theta)$ on $|\phi\rangle$ is well defined.
This is, however, straightforward, because the representation of the algebra~(\ref{w1})
can be explicitly constructed using~(\ref{w2}) and well known Fock--space technique.

Finally consider the states saturating (\ref{w12a}). Inspired by the standard Fock space 
construction and the form of the commutation rule~(\ref{w1a}) we define creation/anihilation
operators
\begin{equation}
\begin{array}{ll}
b=\frac{1}{\sqrt{2\theta}}(\hat{x}_1+i\hat{x}_2), & 
b=\frac{1}{\sqrt{2\theta}}(\hat{x}_1-i\hat{x}_2)
\end{array}
\label{w23}
\end{equation}
or, by (\ref{w2}),
\begin{equation}
\label{w24}
\begin{split}
b&=\frac{1}{\sqrt{2\theta}}\left(\tilde{x}_1-\frac{\theta}{2\hbar}\tilde{p}_2+
i\left(\tilde{x}_2+\frac{\theta}{2\hbar}\tilde{p}_1\right)\right)=\\
&=\frac{1}{\sqrt{2\hbar}}\left(
\frac{1}{\sqrt{2}}\left(\sqrt{\frac{2\hbar}{\theta}}\tilde{x}_1-\sqrt{\frac{\theta}{2\hbar}}
\tilde{p}_2\right)+
\frac{i}{\sqrt{2}}\left(\sqrt{\frac{2\hbar}{\theta}}\tilde{x}_2+\sqrt{\frac{\theta}{2\hbar}}
\tilde{p}_1\right)
\right)
\end{split}
\end{equation}
Define the dilatation operator
\begin{equation}
D=\frac{1}{2\hbar}([\tilde{x}_1,\tilde{p}_1]_{+}+[\tilde{x}_2,\tilde{p}_2]_{+})
\label{w25}
\end{equation}
It is simple to check that
\begin{equation}
b=e^{\frac{i}{2}\ln(\frac{2\hbar}{\theta})D}
\left(\frac{1}{\sqrt{2}}\left(
\frac{\tilde{x}_1+i\tilde{x}_2}{\sqrt{2\hbar}}
\right)
+\frac{i}{\sqrt{2}}\left(
\frac{\tilde{p}_1+i\tilde{p}_2}{\sqrt{2\hbar}}
\right)\right)e^{-\frac{i}{2}\ln(\frac{2\hbar}{\theta})D}
\label{w26}
\end{equation}
or
\begin{equation}
b=
e^{\frac{i}{2}\ln(\frac{2\hbar}{\theta})D}
a_{-}
e^{-\frac{i}{2}\ln(\frac{2\hbar}{\theta})D},
\label{w27}
\end{equation}
where
\begin{equation}
a_{-}=\frac{1}{\sqrt{2}}(a_1+ia_2),
\label{w28}
\end{equation}
is the anihilation operator carrying definite angular momentum.
The complementary anihilation operator is 
\begin{equation}
a_{+}=\frac{1}{\sqrt{2}}(a_1-ia_2).
\label{w29}
\end{equation}
The following prescription can be now given for constructing the states saturating (\ref{w12a}).
We construct the general state saturating, say, 
$\Delta\tilde{x}_1\Delta\tilde{p}_1\geq\frac{\hbar}{2}$ and then replace 
$a_1\rightarrow a_{-}$, $a_2\rightarrow a_{+}$; finally,
$\exp(\frac{i}{2}\ln(\frac{2\hbar}{\theta})D)$ is applied. Again,
one can check explicitly that the whole procedure is well defined.

{\ \newline\noindent\Large\bf Acknowledgment\newline\ \newline}
Piotr Kosi\'nski kindly acknowledges numerous discussions with Prof. Pawe\l{} Ma\'slanka.

\end{document}